\documentclass[twocolumn,times]{aastex631}

\usepackage[T1]{fontenc}
\usepackage[utf8]{inputenc} 
\usepackage{amsmath}

\newcommand\ergs{erg~s$^{-1}$}

\shorttitle{VLBI observations of ULXs}
\shortauthors{Wang et al.}

\graphicspath{{./}{fig/}}
\hypersetup{linkcolor=magenta,citecolor=blue,filecolor=cyan,urlcolor=blue}

\begin{document}

\title{Deep VLBI constraints on compact radio cores in four ultraluminous X-ray sources}

\correspondingauthor{Hua Feng}
\email{hfeng@ihep.ac.cn}
\author[0000-0002-7351-5801]{Ailing Wang}
\affiliation{State Key Laboratory of Particle Astrophysics, Institute of High Energy Physics, Chinese Academy of Sciences, Beijing 100049, People's Republic of China}
\affiliation{Spallation Neutron Source Science Center, 523803 Dongguan, China}
\author[0000-0001-7584-6236]{Hua Feng}
\affiliation{State Key Laboratory of Particle Astrophysics, Institute of High Energy Physics, Chinese Academy of Sciences, Beijing 100049, People's Republic of China}
\author[0000-0003-4341-0029]{Tao An}
\affiliation{Department of Astronomy, University of Science and Technology of China, Hefei, Anhui 230026, China}
\affiliation{Shanghai Astronomical Observatory, Chinese Academy of Sciences, 80 Nandan Road, Shanghai 200030, China}

\author[0009-0008-8549-8069]{Yijia Zhang}
\affiliation{Department of Astronomy, Tsinghua University, Beijing 100084, People's Republic of China}
\author[0000-0002-2322-5232]{Jun Yang}
\affiliation{Department of Space, Earth and Environment, Chalmers University of Technology, Onsala Space Observatory, SE-439 92 Onsala, Sweden}
\author[0000-0002-4622-796X]{Roberto Soria}
\affiliation{INAF—Osservatorio Astrofisico di Torino, Strada Osservatorio 20, I-10025 Pino Torinese, Italy}
\affiliation{Sydney Institute for Astronomy, School of Physics A28, The University of Sydney, Sydney NSW 2006, Australia}
\author[0000-0002-2705-4338]{Lian Tao}
\affiliation{State Key Laboratory of Particle Astrophysics, Institute of High Energy Physics, Chinese Academy of Sciences, Beijing 100049, People's Republic of China}
\author[0000-0002-7930-2276]{Thomas Russell}
\affiliation{INAF, Istituto di Astrofisica Spaziale e Fisica Cosmica, Via U. La Malfa 153, I-90146 Palermo, Italy}
\author[0000-0001-9346-3677]{Jing Guo}
\affiliation{Department of Astronomy, Tsinghua University, Beijing 100084, People's Republic of China}
\author[0000-0003-4498-9925]{Liang Zhang}
\affiliation{State Key Laboratory of Particle Astrophysics, Institute of High Energy Physics, Chinese Academy of Sciences, Beijing 100049, People's Republic of China}

\begin{abstract}

We present high-sensitivity Very Long Baseline Interferometry (VLBI) observations of four ultraluminous X-ray sources (ULXs): Holmberg II X-1, IC 342 X-1, NGC 6946 X-1, and NGC 925 X-1. No compact emission was detected on milliarcsecond scales, with rms noise levels reaching approximately 5--20 $\mu$Jy. 
The corresponding $5\sigma$ flux density upper limits reach $\sim 26\,\mu\mathrm{Jy}$, implying radio luminosity limits $L_{\rm R} \lesssim 2 \times 10^{33}\,\mathrm{erg\,s^{-1}}$.
This disfavors any persistently bright hard-state-like compact core at our sensitivity level. 
The previously reported VLBI core in Holmberg II X-1 exhibits significant long-term variability, broadly consistent with an overall decline over the past decades. This behavior is consistent with emission from optically-thin ejecta undergoing adiabatic expansion. 
The VLBI non-detections may reflect intrinsically weak/intermittent compact emission, and/or low--surface--brightness structure that is resolved out by VLBI, and/or absorption/propagation effects such as free--free absorption in dense, ionized winds.

\end{abstract}


\section{Introduction}
\label{sec:intro}

Recent observations of ultraluminous X-ray sources (ULXs) suggest that the majority of them are powered by supercritical accretion onto stellar mass compact objects \citep[for reviews see][]{Kaaret2017, King2023, Pinto2023}, a physical regime that is not well understood. 
Numerical simulations involving general relativity have revealed that, besides the massive winds launched due to radiation pressure, supercritical accretion may also drive relativistic jets along the symmetric axis of the accretion flow \citep{sadowski2015, Asahina2024, Zhang2026}. 
However, the detailed mechanism of jet launch is still controversial. 

Observationally, SS~433 remains the most proximate and thoroughly documented archetype of a compact object undergoing supercritical accretion, distinguished by its well-characterized relativistic ($v \approx 0.26 c$) jets \citep{Fabrika2004,2004ApJ...616L.159B,2016MNRAS.461..312J}.
It displays a pair of precessing, relativistic jets with a flux density of about $\sim$1~Jy at the core (at the arcsecond scale) and a mechanical power of about $10^{39}$~\ergs. 
Its binary core has an X-ray luminosity of only $\sim$10$^{34-35}$ erg s$^{-1}$ \citep{2005MNRAS.358..860M}, but that is because we are looking at the system almost edge-on, and most of the X-ray photons from the central engine are absorbed or blocked by the thick disk and outflows, along our line of sight. If we looked at SS~433 pole-on, it would likely appear to us as a ULX \citep{2021MNRAS.506.1045M,2021AstBu..76....6F}. 
The jets, after interacting with the interstellar medium, power the surrounding radio nebula W50, which has a total flux density of 71~Jy at 1.4~GHz and 
a physical size of $\approx$100 $\times$ 200 pc \citep{Dubner1998} at a distance of 5 kpc \citep{Su2018}.

This raises fundamental questions in accretion physics: whether supercritical accretion can drive collimated, relativistic jets in addition to broad disk outflows; in what situations jets and outflows co-exist.
Addressing these requires radio observations of ULXs to search for signatures of jets.
In a number of well-studied ULXs, radio observations have revealed optically thin, extended radio emission around them, spatially coincident with the optical emission-line nebula, due to shock-ionization by outflows from the central source or photoionization in a few cases \citep{VanDyk1994, 1997ApJS..109..417L, 2005ApJ...623L.109M, Lang2007, 2012ApJ...749...17C, Urquhart2018, Berghea2020, Soria2021, 2024MNRAS.534..645B}.
However, it is unclear whether the outflow is due to relativistic jets or wide-angle winds. 
NGC 7793 S26 shows a pair of lobes seen in X-ray, optical, and radio, with a morphology similar to that of W50, seemingly in favor of the jet scenario \citep{Pannuti2002, Pakull2010, Soria2010}. 
Another SS~433-like system is the microquasar NGC~300~S10, which shows strong evidence for a collimated jet with an estimated power of $\sim 10^{39}\,\mathrm{erg\,s^{-1}}$, while its apparent X-ray luminosity is relatively low, likely because the central emission is absorbed/scattered for a near edge-on geometry \citep{2019MNRAS.482.2389U}.
Holmberg II X-1 displays a triple radio structure \citep{2014MNRAS.439L...1C}, with the central component being optically thin, resolved (0.26~pc) and variable \citep{Cseh2015}, reminiscent of collimated jets. 
Very long baseline interferometry (VLBI) observations of ULXs have identified unresolved compact radio cores in a few cases \citep{Mezcua2011, Mezcua2013}.
The high radio flux of those cores suggests an alternative interpretation of those sources as sub-Eddington (low/hard state) intermediate mass black holes (IMBHs) or low-mass AGN according to the fundamental plane relations \citep{Mezcua2013}. 
Sporadic or episodic radio activity has been most clearly reported in the transient ULX in M31 \citep{Middleton2013} and the IMBH candidate ESO 243-49 HLX-1 \citep{2012Sci...337..554W, Cseh2015}.

Therefore, it is highly interesting to perform high-sensitivity radio observations of ULXs with VLBI techniques. 
Detection of compact radio cores will be strong evidence in favor of the jet scenario and may help us distinguish between the observational properties of jet cores in the low/hard state and in the super-Eddington regime \citep{Merloni2003}.
In this work, we report VLBI observations of four ULXs (\S~\ref{sec:targets}). The observations, data reduction and results are elaborated in \S~\ref{sec:obs-result},
and discussed in \S~\ref{sec:discussion}. Brief conclusions are presented in \S~\ref{sec:conclusion}. 

\section{Targets}
\label{sec:targets}

We observed Holmberg II X-1, IC 342 X-1, NGC 6946 X-1, and NGC 925 X-1, which are ULXs with a known radio counterpart. These sources have previously been detected at radio wavelengths with the Karl G. Jansky Very Large Array (VLA) or EVN, including reports of compact core candidates or clearly resolved radio structures or nebulae. They therefore represent the most suitable targets for deeper, higher-resolution VLBI observations aimed at testing the presence of compact jet cores on milliarcsecond (mas) scales.

\subsection{Holmberg II X-1}

Holmberg II X-1 \citep[$d=3.1$ \,Mpc;][]{2000ApJS..128..431F} 
is surrounded by an optical emission-line nebula powered by photoionization \citep{Pakull2002, 2004MNRAS.351L..83K, 2005A&A...431..847L}. 
An extended (tens of pc) radio nebula with a steep spectrum was found to be spatially  consistent with the optical nebula \citep{2005ApJ...623L.109M,2012ApJ...749...17C}. VLA A-configuration imaging resolved the arcsecond-scale emission into a triple morphology consisting of a central component plus two fainter outer components aligned roughly along the same axis \citep{2014MNRAS.439L...1C}. A quasi-simultaneous EVN and VLA campaign detected faint, marginally resolved emission at 1.6\,GHz but no 5\,GHz core down to tens of $\mu$Jy \citep{2015MNRAS.452...24C}, favoring optically-thin synchrotron emission from discrete ejecta (or hotspots) rather than a persistent, compact, self-absorbed flat-spectrum jet base.

\subsection{IC 342 X-1}

IC 342 X-1 \citep[$d = 3.3$\,Mpc;][]{2002AJ....124..839S} is associated with a shock-ionized optical ``tooth'' nebula \citep{Pakull2002, 2003MNRAS.342..709R}, and also an extended radio counterpart interpreted as optically thin synchrotron emission \citep{2012ApJ...749...17C}. At the IC 342 X-1 position, the radio flux density is $\sim90\,\mu$Jy, but this is likely contaminated or even dominated by the nebula emission.
Previous EVN observations at 1.6\,GHz yielded a non-detection of any mas-scale core ($<35~\mu$Jy at $5\sigma$; \citealt{2012ApJ...749...17C}).
Subsequent VLA monitoring likewise did not detect any emission at a similar sensitivity \citep{2014MNRAS.444..642M}. 

\subsection{NGC 6946 X-1}

NGC 6946 X-1 ($d=7.7$\,Mpc; \citealt{2018AJ....156..105A}) resides in the optical nebula MF16, which is argued to be powered by both shock and photoionization \citep{2001AJ....121.1497B, Abolmasov2008}.
Arcsecond-resolution VLA imaging finds mJy-level flux densities at GHz frequencies, with a steep spectrum and an extended morphology coincident with the optical nebula \citep{1997ApJS..109..417L,2024MNRAS.534..645B}.

\subsection{NGC 925 X-1}
NGC 925 X-1 ($d=9.3$\,Mpc; \citealt{Saha2006}) is spatially coincident with an optical emission line nebula possibly due to X-ray irradiation \citep{Heida2016, 2021ApJ...906...42L}. 
Recently, a radio counterpart was reported by matching the ULX position with Very Large Array Sky Survey (VLASS) and the Rapid ASKAP Continuum Survey (RACS) images \citep{Zhang2026}.
The radio emission is unresolved with the survey data, showing a flux density of 0.6\;mJy at 3\;GHz with a steep spectral index of $-1.1$.

In sum, previous observations have shown evidence that the four ULXs may have powerful jets or winds, being good candidates for our VLBI observations. 

\section{Data Reduction and Results}
\label{sec:obs-result}

\begin{deluxetable*}{cccccccc}
\centering
\tablecaption{VLBA and EVN observation logs.}
\colnumbers 
\label{tab:VLBA_log}
\tablehead{
\colhead{Obs.\ date} & \colhead{Source} & \colhead{Freq.} & \colhead{code} & \colhead{$B_{\rm maj}\times B_{\rm min}$, $B_{\rm PA}$} & \colhead{rms} & \colhead{$\log \frac{L_{\rm R}}{\rm erg\,s^{-1}}$} & \colhead{Calibrator} \\
\colhead{} & \colhead{} & \colhead{(GHz)} & \colhead{} & \colhead{(mas$\times$mas, \degr)} & \colhead{($\mu$Jy beam$^{-1}$)} & \colhead{} & \colhead{}
}
\startdata
        2024-11-26   & Holmberg II X-1 & 6.2     &BF135AC   & 3.0 $\times$ 1.7, $-21^{\circ}$                & 10                  & $<33.55$    &J0831+7035    \\
        2024-12-02   & Holmberg II X-1 & 6.2     &BF135BC   & 2.5 $\times$ 1.3, $-18^{\circ}$                & 9.1                 & $<33.51$    &J0831+7035    \\ 
        2024-12-06   & Holmberg II X-1 & 6.2     &BF135CC   & 2.5 $\times$ 1.3, $-18^{\circ}$                & 8.9                 & $<33.50$    &J0831+7035    \\
        2024-12-09   & Holmberg II X-1 & 6.2     &BF135DC   & 2.5 $\times$ 1.3, $-20^{\circ}$                & 8.8                 & $<33.50$    &J0831+7035    \\
           Stack    & Holmberg II X-1 & 6.2     &BF135C    & 2.5 $\times$ 1.3, $-19^{\circ}$                 & 5.2                 & $<33.27$    &J0831+7035    \\
        2024-12-18   & Holmberg II X-1 & 1.6     &BF135AL   & 9.7 $\times$ 5.1, $-8.0^{\circ}$               & 17                  & $<33.19$    &J0831+7035    \\
        2025-01-02   & Holmberg II X-1 & 1.6     &BF135BL   & 9.8 $\times$ 5.0, $-22^{\circ}$                & 19                  & $<33.24$    &J0831+7035    \\
        2025-01-08   & Holmberg II X-1 & 1.6     &BF135CL   & 9.9 $\times$ 5.0, $-22^{\circ}$                & 17                  & $<33.19$    &J0831+7035    \\
        2025-01-10   & Holmberg II X-1 & 1.6     &BF135DL   & 8.7 $\times$ 4.7, $-16^{\circ}$                & 25                  & $<33.36$    &J0831+7035    \\ 
           Stack     & Holmberg II X-1 & 1.6     &BF135L    & 9.5 $\times$ 5.1, $-16^{\circ}$                & 8.6                 & $<32.90$    &J0831+7035    \\
        2025-02-11   &IC 342 X-1       & 1.6     &BF137L1   & 9.6 $\times$ 5.0, $-8.4^{\circ}$               & 18                  & $<33.27$    &J0344+6827    \\
        2025-02-14   &IC 342 X-1       & 6.2     &BF137C1   & 3.8 $\times$ 2.8, $-20^{\circ}$                & 13                  & $<33.72$    &J0344+6827    \\
        2025-03-22   &NGC 6946 X-1     & 1.6     &BF137L4   & 11 $\times$ 5.6, $-16^{\circ}$                 & 19                  & $<34.03$    &J2035+5821    \\
        2025-03-21   &NGC 6946 X-1     & 6.2     &BF137C4   & 2.6  $\times$ 1.3, $-8.5$                      & 10                  & $<34.34$    &J2035+5821    \\
        2025-04-09   &NGC 925 X-1      & 1.7     &EZ035     & 6.2 $\times$ 2.2, 3.9                          & 17                  & $<34.18$    &J0226+3421    \\
        2025-08-11   &NGC 925 X-1      & 1.6     &BZ116     & 14 $\times$ 4.6, $-$6.1                        & 21                  & $<34.24$    &J0226+3421    \\ 
		\hline
	\enddata
\tablecomments{
Columns: (1) observation date, (2) source name, (3) frequency in GHz, (4) project code, (5) beam size and position angle, (6) off-source image rms noise measured in a nearby source-free region, (7) 5$\sigma$ luminosity upper limit, and (8) calibrator. \\
Overall, the VLBA observations were carried out with the standard array of BR, FD, HN, KP, LA, MK, NL, OV, PT, and SC, with KP missing in BF135DL, MK missing in BF137C1, SC missing in BF137L4, and BR missing in BZ116, while all other epochs used a complete or nearly complete VLBA antenna configuration. There are 8 stations in EVN observation, including Jb, Wb, Ef,  O8,  T6,  Tr,  Hh, and Ir.
}
\end{deluxetable*}

We observed Holmberg II X-1, IC 342 X-1, and NGC 6946 X-1 with the National Radio Astronomy Observatory (NRAO) Very Long Baseline Array (VLBA) in 2024--2025 (programs BF135, BF137; Table~\ref{tab:VLBA_log}). Each session lasted 6\,hr and was conducted at L band (central frequency $\simeq1.6$\,GHz) and C band (central frequency $\simeq6.2$\,GHz). We observed NGC 925 X-1 at L band with both the European VLBI Network (EVN; program EZ035) and the VLBA (program BZ116; Table~\ref{tab:VLBA_log}).  The VLBA observations used dual circular polarization with 2-bit sampling at a recording rate of 4 gigabits per second (Gbps), configured as four 128\,MHz intermediate frequencies (IFs) (total bandwidth 512\,MHz per polarization). 
The EVN observations were conducted with 8 antennas, using four IFs with 32 MHz bandwidth each, four polarizations, 64 spectral channels per subband, and 2-second integrations.

Because ULXs are intrinsically faint at mas scales, we employed phase referencing using a 5-min nodding cycle (1\,min on the phase calibrator and 4\,minutes on target). The phase calibrators are listed in Table~\ref{tab:VLBA_log}. Most epochs used the full 10-antenna VLBA array, with the few antenna dropouts noted in Table~\ref{tab:VLBA_log}.

Correlation was performed with Distributed FX (DiFX) software correlator at the VLBA correlator in Socorro \citep{2011PASP..123..275D} and EVN software correlator SFXC at Joint Institute for VLBI ERIC (JIVE) \citep{2015ExA....39..259K}, which minimizes time and bandwidth smearing over the region imaged around each ULX. The correlated data were transferred to the China Square Kilometre Array Regional Centre \citep{2019NatAs...3.1030A,2022SCPMA..6529501A} for calibration and imaging.

VLBA calibration followed standard VLBA procedures in Astronomical Image Processing System \citep[\textsc{AIPS}, ][]{2003ASSL..285..109G}. We applied a priori amplitude calibration using measured system temperatures and gain curves (\textit{ANTAB} and \textit{APCAL}). Dispersive ionospheric delays were corrected with \textit{TECOR} using total electron content maps from the NASA CDDIS archive, and Earth-orientation parameters were updated using \textit{CLCOR}. Parallactic-angle corrections were applied with \textit{VLBAPANG}. Instrumental phase and delay offsets between IFs were corrected using pulse-cal signals (\textit{VLBAMPCL}), and residual multi-band delays and rates were solved by global fringe fitting on the phase calibrators (\textit{FRING}). We also derived complex bandpass solutions (\textit{BPASS}) and applied them before imaging.

EVN calibration was performed in AIPS following standard EVN VLBI procedures. Instrumental phase and delay offsets between IFs were corrected using fringe fitting on strong calibrators. Residual multi-band delays, delay rates, and time-dependent phases were then solved by global fringe fitting on the phase calibrators, with the solutions transferred to the target source. Complex bandpass responses were derived from a dedicated bandpass calibrator and applied to the data (\textit{BPASS}). Where necessary, instrumental delay and fringe-fitting steps were repeated after bandpass calibration to remove any residual subband offsets. The fully calibrated data were subsequently used for imaging and further analysis.

We imaged the phase calibrators in DIFMAP \citep{1994BAAS...26..987S} and iterated CLEAN and self-calibration to refine calibrator structure models; the resulting solutions were transferred back to AIPS and applied to the target datasets. Targets were imaged in Stokes~I using natural weighting to maximize point-source sensitivity, and we additionally generated $(u, v)$-tapered images to search for partially resolved emission. Because none of the ULXs is detected, no self-calibration was performed on the targets. 

For Holmberg II X-1, we concatenated the four epochs at each band using DBCON to improve image sensitivity. The final synthesized beams and rms noise levels are listed in Table~\ref{tab:VLBA_log}.

We imaged each epoch at each frequency using natural weighting to maximize point-source sensitivity and also produced a set of $(u, v)$-tapered images to increase sensitivity to low-surface-brightness emission on  largest VLBA-resolved scales. 
No compact emission is detected at the ULX positions, nor elsewhere within the imaged fields, at either frequency in any epoch. 
We adopt point-source upper limits of $S_{\nu,{\rm lim}} = 5\sigma_{\rm rms}$, where $\sigma_{\rm rms}$ is measured in a nearby source-free region.
The corresponding upper limits in radio luminosity are also computed as $L_{\rm R}\equiv \nu L_\nu$ assuming the source distance, also listed in Table~\ref{tab:VLBA_log}.

\begin{deluxetable*}{cccccc}
\centering
\tablecaption{Radio flux densities and upper limits of the Holmberg II X-1.}
\colnumbers 
\label{tab:history}
\tablehead{
        \colhead{telescope} & \colhead{beam size} & \colhead{Obs. date} & \colhead{Freq.} & \colhead{flux density } & \colhead{Ref.} \\
        \colhead{}          & \colhead{(arcsec $\times$ arcsec)} & \colhead{}  & \colhead{(GHz)}      & \colhead{(mJy)} & \colhead{}
}
\startdata    
        VLA        &1.7 $\times$ 1.5         & 1988,1989,1990,1991 & 4.9 &0.677$\pm$ 0.207        & 1 \\
        VLA        &1.9 $\times$ 1.5         & 1994-04-03   & 1.4     & 1.174 $\pm$ 0.085         & 1 \\
        VLA        &1.5 $\times$ 1.0         & 2007-12-08   & 4.8     & 0.114$\pm$0.014           & 2 \\
        VLA        &2.4 $\times$ 1.8         & 2008-04-21   & 8.5     & 0.145$\pm$0.015           & 2 \\
        VLA        &0.41 $\times$ 0.30       & 2012-12-16/17& 5.0     & 0.151$\pm$0.003           & 3 \\
        EVN        &0.012 $\times$ 0.010     & 2014-01-14   & 1.6     & 0.053$\pm$0.015           & 4 \\
        EVN        &0.003 $\times$ 0.002     & 2014-04-28   & 5.0     & $<0.060$                  & 4 \\
        VLA        &0.41 $\times$ 0.22       & 2014-05-25   & 9.0     & 0.015$\pm$0.007           & 4 \\
        VLASS      &3.4 $\times$ 2.0         & 2017-09-18   & 3.0     & $<0.803$                  & 5 \\
        VLASS      &3.3 $\times$ 2.1         & 2020-09-10   & 3.0     & $<0.778$                  & 5 \\
        VLASS      &3.4 $\times$ 2.3         & 2023-02-17   & 3.0     & $<0.605$                  & 5 \\
        VLBA       &0.0025 $\times$ 0.0013   & 2024-11-26 to 2024-12-09  & 6.2 & $<0.026$                & this work \\
        VLBA       &0.0095 $\times$ 0.0051   & 2024-12-18 to 2025-01-10 & 1.6 & $<0.043$                  & this work \\
        VLASS      &3.2 $\times$ 2.0         & 2026-01-02   & 3.0     & $<0.640$                  & 5 \\
	\enddata
\tablecomments{
 Columns: (1) telescope, (2) beam size, (3) observing date, (4) frequency, (5) flux density or 5$\sigma$ upper limit, and (6) references.
\tablerefs{
1.\ \citet{2005ApJ...623L.109M}; 
2.\ \citet{2012ApJ...749...17C}; 
3.\ \citet{2014MNRAS.439L...1C}; 
4.\ \citet{2015MNRAS.452...24C}; 
5.\  \citet{2020PASP..132c5001L}.
}
}
\end{deluxetable*}

\section{Discussion}
\label{sec:discussion}

Our VLBI observations yielded non-detections of the four ULXs, allowing us to place stringent upper limits (down to several tens of $\mu$Jy at 5$\sigma$) on any compact mas-scale core emission, and more generally on any mas-scale jet/lobe-like structure with sufficiently high surface brightness. More extended, low–surface-brightness emission on sub-arcsecond/arcsecond scales would be resolved out by VLBI and is not constrained by these limits.
Here we discuss the possible physical nature in line with the non-detections.

\subsection{Comparison with X-ray binaries}

Adopting the measured flux upper limits, the beam-averaged brightness-temperature limits can be expressed as
\begin{equation}
T_{\rm b} < 1.22\times10^{12}
\left(\frac{S_{\nu, {\rm lim}}}{\rm Jy}\right)
\left(\frac{\nu}{\rm GHz}\right)^{-2}
\left(\frac{\theta_{\rm maj}\theta_{\rm min}}{\rm mas^2}\right)^{-1}\ {\rm K} \, ,
\end{equation}
where $\theta_{\rm maj}$ and $\theta_{\rm min}$ are the beam size along the major and minor axis, respectively.
Given the deepest stacked observations of Holmberg II X-1 (Table~\ref{tab:VLBA_log}) as an example, one obtains $T_{\rm b} < 3 \times 10^5$\,K at 6.2\,GHz and $T_{\rm b} < 4 \times 10^5$\,K at 1.6\,GHz. 
These values are lower than the brightness temperatures typically reported for bright hard-state compact cores in black-hole X-ray binaries \citep[e.g., $T_{\rm B} > 10^6$\;K;][]{2000ApJ...543..373D, 2001MNRAS.327.1273S, 2009ApJ...706L.230M, 2013MNRAS.432.1319P}, noting that our limits are beam-averaged and thus primarily constrain the surface brightness on mas scales.

These limits primarily constrain the mas-scale surface brightness. They therefore disfavor any persistently bright, Doppler-boosted compact core at our achieved sensitivity, but they do not by themselves exclude intrinsically high-$T_{\rm b}$ emission confined to angular scales $\ll$ the synthesized beam if the total flux density is below our detection thresholds, nor do they rule out low--surface-brightness structure distributed on larger scales that would be resolved out by VLBI. Consequently, the present non-detections indicate that none of the four ULXs shows a stable, bright VLBI core during the observed epochs, rather than uniquely requiring a complete absence of compact jet activity.

If SS~433 were placed at 3.1~Mpc, a 2.5~mas VLBA beam would correspond to a physical scale of $\approx 0.038$~pc, equivalent to $\approx 1.6^{\prime\prime}$ at the distance of SS~433 ($\sim$5~kpc). This is comparable to the extent of the central core region detected with the VLA with a 5~GHz radio luminosity $\nu L_{\nu} \sim 10^{32}$~erg~s$^{-1}$ \citep{2004MNRAS.354.1239S}. Its core flux density at 3.1~Mpc would drop to the $\sim$2 $\mu$Jy. 
Our VLBA limits therefore do not exclude SS~433-like jets if their compact radio cores are as faint at GHz frequencies as implied by the SS 433 scaling.
However, we note that the outflow mechanical power estimated from shock-ionized nebulae in ULXs is generally one or two orders of magnitude higher than that in SS~433 \citep{Pakull2002}.

\subsection{Temporal variability}

The EVN observation on 2014-Jan-14 detected a flux density of $53 \pm 15\; \mu$Jy at 1.6 GHz from Holmberg II X-1 \citep{2015MNRAS.452...24C}, while our $3\sigma$ upper limit is 25.8\;$\mu$Jy and  $5\sigma$ upper limit is 43\;$\mu$Jy at the same band with similar beam sizes.
The two measurements are inconsistent with each other at a significance greater than 3$\sigma$, suggesting that the radio emission is transient or variable at the VLBI scale. 
In particular, \citet{2015MNRAS.452...24C} revealed that the radio core is extended on mas scale with unboosted emission, and faded by a factor of 7.3 or more over a course of 1.5 years, eventually falling below the detection threshold in later epochs, compatible with the scenario of adiabatic expansion.
Along with the steep spectrum \citep[$\alpha = -0.8$;][]{2014MNRAS.439L...1C}, the central core is more likely an ejecta rather than a compact steady jet.

\begin{figure}
    \centering
    \includegraphics[width=1\linewidth]{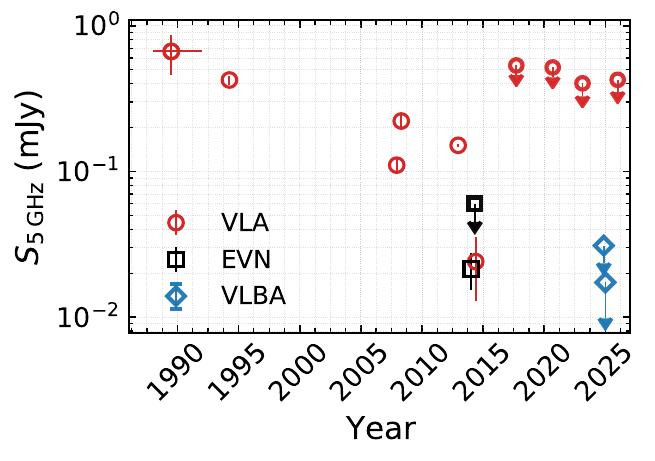}
    \caption{
    Radio light curve of Holmberg~II~X--1 at 5 GHz. 
    Compilation of published radio measurements and upper limits for the central radio region of Holmberg II X-1, scaled to 5 GHz assuming a power-law spectral index $\alpha = -0.8$ \citep{Cseh2014} for illustrative comparison.
    Red circles show VLA measurements, black squares show EVN measurements, and blue diamonds show VLBA measurements (table \ref{tab:history}). Downward triangles denote upper limits for non-detections. The early VLA measurements obtained between 1988 and 1991 are shown as a single point at the midpoint epoch (1989.5), with the horizontal error bar indicating the full time span of the observations. 
    The compiled measurements span a wide range of angular resolutions; different data points likely include different amounts of extended (non-core) emission (table \ref{tab:history}).
    }
    \label{fig:HoII-lc}
\end{figure}

We compiled published VLA, EVN, VLBA, and VLASS measurements for the central radio region of Holmberg II X-1 in Table \ref{tab:history} and show them in Figure \ref{fig:HoII-lc} after scaling them to 5 GHz assuming $\alpha = -0.8$ for illustrative comparison. We emphasize, however, that this spectral index was inferred from a limited set of earlier measurements and should not be regarded as a time-independent property of the source. Given the known variability of Holmberg II X-1, the spectral index may also vary with epoch and source state, and free–free absorption at lower frequencies could further affect the spectral shape and hence the extrapolated fluxes. We also caution that the compiled measurements span a wide range of angular resolutions, from mas-scale EVN/VLBA data to sub-arcsecond and arcsecond-scale VLA/VLASS observations. The lower-resolution measurements may therefore include significant contributions from expanding ejecta and/or emission from the surrounding ionized nebula,
rather than isolating the compact core alone. Taken together, the available data are broadly consistent with long-term variability, and possibly fading central radio emission, but they do not uniquely establish a monotonic decline of an isolated compact core. This ambiguity further strengthens the case for sensitive VLBI monitoring of ULXs, which is essential for disentangling any genuine compact core from nearby ejecta/bolides and more extended nebular emission.

Sporadic or episodic radio activity has been seen in two ULXs, a transient ULX in M31 \citep{Middleton2013} and ESO 243-49 HLX-1 \citep{2012Sci...337..554W, Cseh2015}.
However, these two ULXs show distinct X-ray behaviors from canonical ULXs like the four in this work. 
Most ULXs have been active since they were discovered in the 1980s with the Einstein telescope, while the M31 transient was in the outburst for less than a year \citep{Middleton2013} and HLX-1 has been in quiescence after several cycles of outbursts (or large amplitude variation) since its discovery \citep{2020MNRAS.491.5682L}. 
HLX-1 shows intermittent, short-lived transient radio detections during X-ray outbursts, especially around the hard-to-soft state transition. 
ATCA detected tens of $\mu$Jy emission at 5 and 9 GHz that disappeared in adjacent epochs, and the strong variability favors discrete jet ejection events rather than a steady core \citep{2012Sci...337..554W,Cseh2015}.
These ejecta are expected when an X-ray binary transitions from the hard to soft state \citep{2004MNRAS.355.1105F,2009MNRAS.396.1370F}, while other ULXs are not staying at any of these sub-Eddington spectral states \citep{2015MNRAS.454.3134M}. 

\subsection{Absorption in the close vicinity}

The majority of ULXs are high mass X-ray binaries at the high-luminosity end \citep{2012MNRAS.419.2095M}.
Population studies suggest that they may contain an evolved companion \citep{2008ApJ...688.1235M}.
A B9Ia supergiant was identified in NGC 7793 P13 \citep{2014Natur.514..198M}, and a helium star was identified in NGC 247 X-1 \citep{Zhou2023a}.
Also, massive disk winds are believed to be ubiquitous in supercritical accretion, as suggested by theory \citep{1982ApJ...256..681M, Poutanen2007} and supported by observations \citep{2015NatPh..11..551F, 2019ApJ...871..115Z, 2021ApJ...906...36Q}.  
Thus, ULXs may be embedded in an environment filled with stellar and/or disk winds.

For a spherically symmetric ionized wind with mass-loss rate $\dot{M}$ and terminal speed $v$, the electron density scales as $n_e(r)\propto \dot{M}\,r^{-2}v^{-1}$, and the emission measure scales as ${\rm EM}\propto \dot{M}^2 v^{-2} R_0^{-3}$, where $R_0$ is the launching radius of the wind. 
Using the standard approximation, 
\begin{equation}
\tau_{\rm ff} \simeq 0.082\, \left(\frac{T_e}{{\rm K}}\right)^{-1.35}
\left(\frac{\nu}{\rm GHz}\right)^{-2.1}
\left(\frac{{\rm EM}}{{\rm pc\,cm^{-6}}}\right),
\end{equation}
and adopting the representative wind velocity and temperature for ULXs detected with ionized winds \citep{Pinto2016}, one obtains
\begin{multline}
\tau_{\rm ff}(6\,{\rm GHz}) \approx 
0.53\,
\left(\frac{\dot{M}}{10^{-4}\,M_\odot\,{\rm yr^{-1}}}\right)^2
\left(\frac{v}{0.1c}\right)^{-2} \\
\times
\left(\frac{T_e}{10^7\,{\rm K}}\right)^{-1.35}
\left(\frac{R_0}{10^{13}\,{\rm cm}}\right)^{-3}.
\end{multline}

Here we assume that ULXs have a mass-loss rate comparable to that of SS~433 \citep{Fabrika2004} and a launching radius (i.e., the inner boundary of the absorbing screen) comparable to the binary separation. Under these representative parameters, free--free optical depths of order unity at 6~GHz (and substantially larger at 1.6~GHz) are possible for supercritical mass-loss rates and sufficiently small $R_0$. In that case, free--free absorption could contribute to suppressing the apparent GHz-band VLBI core flux (in particular for the cases of IC~342~X-1, NGC~6946~X-1, and NGC~925~X-1), although the present non-detections can equally be explained by intrinsically weak/intermittent compact emission and/or low--surface--brightness structure that is resolved out by VLBI. High-frequency VLBI ($\gtrsim 15$~GHz), where $\tau_{\rm ff}$ is expected to decrease rapidly with frequency, would provide an effective test of this scenario.
Such observations, if yielding detections, would also help place useful constraints on the circum-source environment of ULXs and, consequently, on their accretion physics and/or evolutionary history.

\subsection{Fundamental plane and IMBH interpretation}

The radio upper limits derived in this work cannot exclude the possibility that the four ULXs follow the radio-X-ray luminosity correlation found for hard-state X-ray binaries \citep{Corbel2013, 2019ApJ...871...26K}.
If a ULX were a sub-Eddington, hard-state accretor with a scale-invariant compact jet, its core radio luminosity should follow
\begin{equation}
\log L_{\rm R} = 0.60\,\log L_{\rm X} + 0.78\,\log M_{\rm BH} + 7.33,
\end{equation}
where $L_{\rm R}\equiv \nu L_\nu$ at 5\,GHz and $L_{\rm X}$ is the 2--10\,keV luminosity \citep{Merloni2003}. 
Under this assumption, our VLBA limits can be translated into upper limits on $M_{\rm BH}$. Approximating our 6.2\,GHz limits as 5\,GHz limits for a flat core spectrum, we obtain
$M_{\rm BH} \lesssim 3.1\times10^2 L_{\rm X,40}^{-0.77} M_\odot$ (Holmberg II X-1), 
$M_{\rm BH} \lesssim 1.2\times10^3 L_{\rm X,40}^{-0.77} M_\odot$ (IC 342 X-1), 
$M_{\rm BH} \lesssim 7.2\times10^3  L_{\rm X,40}^{-0.77} M_\odot$ (NGC 6946 X-1), 
where $L_{\rm X,40}$ is $L_{\rm X}/(10^{40}\,{\rm erg\,s^{-1})} $.
For context, the fundamental plane based constraints reported for ULXs typically fall in the IMBH regime, with representative values ranging from $M_{\rm BH}\lesssim(1.0\pm0.3)\times10^{3}\,M_\odot$ \citep[IC 342 X-1, ][]{2012ApJ...749...17C} to $\log(M_{\rm BH}/M_\odot)<4.2$--5.9 for several systems \citep{Mezcua2013}, whereas HLX-1 allows $M_{\rm BH}\lesssim2.8\times10^{6}\,M_\odot$ \citep{2014MNRAS.439L...1C} and M82~X-1 has $\sim2.7\times10^{3}\,M_\odot$ with large uncertainties \citep{2025MNRAS.540..239W}.
These limits should be interpreted with caution because ULXs usually occupy the ultraluminous regime rather than canonical hard states \citep[e.g.,][]{2009MNRAS.397.1836G}, where the fundamental plane has not been tested. 
The inflow–outflow coupling and accretion geometry in ULXs may differ substantially from those of the systems on which the fundamental plane is based.
Plus, the radio and X-ray luminosities were not measured simultaneously, which introduces additional uncertainties. 
However, any future secure radio detection of these sources at VLBI resolution could imply the presence of an IMBH \citep[e.g., ][]{2024ApJ...977..211P}.

\subsection{Super-Eddington feedback as a generic astrophysical process}

Beyond ULXs, the central question probed here, \textit{how supercritical accretion partitions energy between radiation, winds, and jets}, is of broad relevance to compact-object feedback across the mass scale. ULXs are among the best nearby laboratories for the physics that is invoked in models of rapid black-hole growth and feedback in dense environments. Our results strengthen a picture in which the time-averaged mechanical impact of supercritical accretion can be large, while the instantaneous compact jet core may be radio weak or hidden. This decoupling has direct implications for interpreting VLBI non-detection of the compact radio component in other high-Eddington systems and for designing future multi-wavelength campaigns aimed at catching the intermittent, potentially transition-linked radio episodes expected if discrete ejecta occur.

\section{Conclusions}
\label{sec:conclusion}

We have performed high-sensitivity VLBI observations of  four ULXs, Holmberg II X-1, IC 342 X-1, NGC 6946 X-1, and NGC 925 X-1, to search for mas-scale radio emission. 

No compact emission is detected in any dataset, with a 5$\sigma$ upper limit as deep as 26~$\mu$Jy at 6.2 GHz for Holmberg II X-1.
For the other sources, the 5$\sigma$ flux-density upper limits are at the level of $\sim$50-100\;$\mu$Jy.
These correspond to core radio-power limits of $L_{\rm R} < (1.6-22) \times 10^{33}$\;\ergs.
These upper limits together with the beam sizes can be translated to brightness temperatures no more than a few times $10^5$\;K in both bands, lower than the beam-averaged brightness temperatures typically inferred for bright hard-state compact cores, while primarily constraining mas-scale surface brightness. 

For Holmberg II X-1, its central radio emission is found to be highly variable.  Overall, it has shown a declining trend over the past 35 years, consistent with the scenario that it is optically thin ejecta undergoing adiabatic expansion. 

Another possibility is that the non-detections are actually due to free-free absorption in the vicinity of the binary system, as the binary system may be immersed in dense, ionized winds.
Future high-frequency VLBI observations may test this scenario and place physical constraints on the accretion physics and/or evolutionary history. 

\begin{acknowledgments}

We thank the referee for the careful reading of the manuscript and for the constructive comments and suggestions, which helped improve the quality and clarity of this work.
HF acknowledges funding support from the National Natural Science Foundation of China under the grant 12025301, the Strategic Priority Research Program of the Chinese Academy of Sciences, and China's  Space Origins Exploration Program.
AW is supported by the National Natural Science Foundation of China under Grant Number 12503018 and the China Postdoctoral Science Foundation under Grant Number of 2025M773197 and 2025T180874.
This work is supported by the National SKA Program of China (grant number 2022SKA0120102). 
TA acknowledges support from Shanghai Oriental Talent Project, the Xinjiang Tianchi Talent Program, the FAST Special Program (NSFC 12041301). 
The VLBI data processing made use of the compute resource of the China SKA Regional Centre \citep{2019NatAs...3.1030A,2022SCPMA..6529501A}. 
The National Radio Astronomy Observatory is a facility of the National Science Foundation operated under cooperative agreement by Associated Universities, Inc. Scientific results from data presented in this publication are derived from the following VLBA project codes: BF135, BF137, BZ116 and EVN project code: EZ035.

\end{acknowledgments}

\vspace{5mm}
\facilities{EVN, VLBA}

\software{astropy\citep{2013A&A...558A..33A,2018AJ....156..123A}, AIPS \citep{2003ASSL..285..109G}, Difmap \citep{1994BAAS...26..987S}}

\bibliography{ulx_vlbi}
\bibliographystyle{aasjournalv7}

\end{document}